\documentclass[a4paper]{jpconf}

\usepackage{graphicx}
\renewcommand{\d}{{\rm d}}

\newcommand{\myskip}[1]{}

 \newcommand{\up}{\uparrow}
  \newcommand{\down}{\downarrow}
 \newcommand{\Up}{\Uparrow}
  \newcommand{\Down}{\Downarrow}

\newcommand{\vA}{{\bf A}}      
\newcommand{\vB}{{\bf B}}      
\newcommand{\vE}{{\bf E}}

\newcommand{\vk}{{\bf k}}   
\newcommand{\vn}{{\bf n}}   
\newcommand{\vv}{{\bf v}}

\newcommand{\vr}{{\bf r}}

\newcommand{\tf}{t_{\rm f}}

\newcommand{\BEQ}{\begin{eqnarray}}      
\newcommand{\EEQ}{\end{eqnarray}}      
\newcommand{\BEA}{\begin{eqnarray}}      
\newcommand{\EEA}{\end{eqnarray}}      
\newcommand{\nn}{\nonumber }      
\renewcommand{\d}{{\rm d}}

\newcommand{\eps}{\varepsilon}      
      
\newcommand{\om}{\omega}

\newcommand{\half}{\frac{1}{2}}

\begin{document}


\title[A subquantum arrow of time]{A subquantum arrow of time\footnote{Published in J. Phys.: Conf. Ser., vol 504 (2014), (1), 012008}}

\author{Theo M. Nieuwenhuizen}

\address { Institute for Theoretical Physics, University of Amsterdam, Science Park 904, P.O. Box 94485, 
 1090 GL  Amsterdam, The Netherlands}

\ead{t.m.nieuwenhuizen@uva.nl }

\begin{abstract}
The outcome of a single quantum experiment is unpredictable, except in a pure-state limit.
The definite process that takes place in the apparatus may either be intrinsically random or be explainable from a deeper theory. 
While the first scenario is the standard lore, the latter implies that quantum mechanics is emergent. 
In that case, it is likely that one has to reconsider radiation by accelerated charges as a physical effect, which thus
must be compensated by an energy input. Stochastic electrodynamics, for example, asserts that the vacuum energy arises from classical
fluctuations with energy $\frac{1}{2}\hbar\omega$ per mode. In such theories the stability of the hydrogen ground state will 
arise from energy input from fluctuations and output by radiation, hence due to an energy throughput. 
That flux of energy constitutes an arrow of time, which we call the ``subquantum arrow of time''. It is related to the stability of matter and it is
 more fundamental than,  e.g.,  the thermodynamic and cosmological arrows.
\end{abstract}



\section{Introduction}

\hfill{\it Time and tide wait for no man }

\hfill{Proverb}

\vspace{3mm}

The arrow of time: why do we get older, never younger? has bothered people and philosophers since long.
The fact that Nature has an arrow of time while being described by reversible dynamical equations such as classical mechanics, electrodynamics, 
quantum mechanics and general relativity, poses a deep problem:  is there an ultimate cause for the arrow of time? 
Of the many answers given, we mention the thermodynamic arrow, relying on the second law (entropy increase),
and the cosmological arrow, which acknowledges that the Universe is expanding while aging. But is there a still more fundamental one?
The aim of the present paper is to investigate whether the {\it emergence of quantum mechanics} \cite{NEmQM12} can provide one.
To get there, we need first to dwell on recent progress on the foundations of quantum mechanics itself. 
 
 \section{On quantum measurement and the interpretation of quantum mechanics}
 
\hfill{\it To understand Nature}

\hfill{ \it we have become accustomed}

\hfill{\it to inconceivable concepts }

\vspace{3mm}

 In textbooks quantum measurement is treated by means of postulates for ideal measurements: the collapse postulate and the Born rule.
To understand this from a more conceptual level, many attempts have been made.
While various approaches stay within standard quantum mechanics, others go beyond it.
 But in the practice of a laboratory a measurement is an interaction between two systems: the tested system and the apparatus.
 This beying remote from postulates,  motivates the study of models aiming to describe realistic ideal measurements.
 The literature on this subject has been  reviewed in  \cite{ABNOpus}. 
 
Not having reached consensus  on the meaning quantum mechanics (QM) after nearly a century 
(as starting point we take the Dreim\"annerarbeit of Born, Jordan and Heisenberg in 1926) 
has had an enormous impact on the philosophical view of the community. 
Instead of admitting that convincing arguments have yet to be filtered out, the case has turned into
``We did our utmost best to find a solution, but there really is none.''
This psychologically understandable but unfortunate position has led to remarkable statements such as: 
``the development from animals to humans has prepared brains incapable to understand the quantum nature of matter'' \footnote{The situation being: 
after we adopt this or that  interpretation,  we run into oddities. A logical conclusion is, however, not to blame the theory but the making of such assumptions.}$^,$
\footnote{A similar statement by Edward Witten about string theory: ``String theory is a part of 21st-century physics that fell by chance into the 20th century'' is
to be regarded as equally remarkable. 
This theory has not reproduced the standard model of particle physics, and even if it would, for what reason would the necessary set of parameters be singled out?
In fact, string theory has  more the looks of a {\it framework}, applicable to different situations, including Fermi liquid theory, than of a fundamental {\it theory of Nature}.}.
We do not have to wonder what Einstein would say about that, he has spent half of his life trying to avoid this trap; so did 't Hooft \cite{tHooftEmQM12}.
Indeed, it is never too late to face the fundamental questions rather than to answer them by ``talking dynamics'' 
(intricate wording instead of convincing, functioning principles).

Since the only point of contact between quantum theory and Nature lies in measurements, 
it seems rather obvious \cite{ABNOpus} that in order to determine the meaning of the theory, 
one must understand what goes on during a measurement.\footnote{That this is not
obvious to everybody, may be seen from several approaches to the interpretation of QM that try to do away with measurements, sometimes
even banning the term. See \cite{Schlosshauer} for a review.} 
Only focussing on the measurement postulates, like most approaches do \cite{Schlosshauer}, 
is a {\it black box} approach, that may be of help for certain questions, but is doomed  to fail
at the most fundamental level. We expect the same to hold for quantum information, in the sense that it can provide us with constraints on
the working of quantum theory, without being able to answer all the fundamental questions that can be answered within QM.

Our basic notion is that an apparatus used to perform a measurement interacts with the tested system, building a special type of interacting quantum systems.
\cite{deMuynck}
The apparatus consists of atoms and the measurement causes a change of its state. On a theoretical level we then have to ask:
how can an ideal measurement be modelled? Obviously, the answer lies in the quantum solid state theory
for two coupled quantum systems, the tested system and the apparatus.
Fortunately, in applications to the solid state the interpretation of quantum mechanics plays a secondary role \cite{Balian_book};
in reverse, the involved one can be expected to have an impact on understanding of measurements.
 
 \subsection{The Curie-Weiss model for a quantum measurement and its dynamics}
 
Considering a model for a quantum measurement, there are three important issues to be explained:
{\it truncation} of the density matrix (vanishing of Schr\"odinger cat terms); 
{\it registration}, the stable indication of the measurement outcome by the pointer variable, and the possibility of {\it reduction},
to update our knowledge about the ensemble of tested systems on the basis of measurement outcomes \cite{ABNOpus}.
For the latter, one needs to explain that single measurements have definite outcomes 
even though quantum mechanics is a wave theory, that is, to be able to speak about individual events, even though quantum mechanics is an ensemble theory.
Either way, one needs to solve  the  {\it measurement problem}.
\footnote{The opinion: 
``the alleged ``measurement problem'' does not exist as a problem of quantum theory''  \cite{Englert}  misses the essence of the
reason for the many interpretations of QM and fails to connect to the answer proposed in ref. \cite{ABNOpus} and the general theory of \cite{ABNOpusculo}. }

 A rather rich but flexible and solvable model is the Curie-Weiss model, introduced in \cite{ABNCW2003} and worked out in detail in \cite{ABNOpus}. 
The system S is a spin $\half$. The aim is to measure its $z$-component by means of an apparatus A, which consists of an Ising magnet M coupled to a bath B.
 The magnet starts in a metastable paramagnetic initial state and will be triggered by the measurement to go to one of its two stable ferromagnetic phases 
 with magnetization  $\pm m_{\rm F}$. Hence the measurement relies on a first order phase transition which amplifies the small quantum signal,
 a situation known from our retina, where a handful of photons can achieve an avalanche of electrons, that sends an electric current to the cortex.
 Not surprisingly, the measurement basis is determined by the interaction Hamiltonian, that is, by the forces exerted by the apparatus on the tested system.
 \footnote{This answers the ``which basis'' question frequently occurring in the measurement literature based on postulates rather than models,
 in particular in those approaches that ``do away with measurements''. }
 
After coupling S to A, there first occurs a {\it dephasing} in the off-diagonal terms of the density matrix, as happens in NMR physics.
Under suitable conditions the now hidden information is lost before recurrences may occur by a {\it decoherence} due to the bath, from which it cannot be recovered. 
Hence a {\it truncation} of off-diagonal elements of the density matrix occurs, commonly called decay of Schr\"odinger cat terms, death of Schr\"odinger cats, and so.
It is a {\it physical process}, due to coupling of S to A, more precisely: first due to the interaction Hamiltonian, then to the bath.
 
In a second stage of the process, the {\it registration} of the result occurs. 
The  magnet M goes from its initial paramagnetic state to the stable up--or--down ferromagnetic state, provided that the initial metastable state has been made
unstable by a strong enough coupling between S and A. This phase transition is exploited to amplify the small quantum signal.

The third stage of the measurement is a new one, not discussed in the literature until recently,
appearing first in version 3 of  the arXiv preprint of our review \cite{ABNOpus}.

 \subsection{The measurement problem: a conundrum for theorists and philosophers}

 \vspace{3mm}
 
 \hfill{\it Win the battle only to lose the war}
 
 \hfill{Proverb}

 \vspace{3mm}
 
 Experience shows that experiments yield definite outcomes. 
 In classical physics this is obvious: when one throws a handful of coins on the ground, each one will expose either head or tail.
 In quantum physics things are not clear, however. 
 Indeed, an unpolarized beam of neutrons can be decomposed as half having spin up and half spin down.
 However, this can be done {\it along any axis}, bringing an infinity of possibilities,
 so the decomposition is a mathematical one, bearing no physics, and, in particular,
 not allowing to suppose that half of the neutrons can be thought to have spin up and the other half spin down along the $z$-axis.\footnote{The combination
of a neutron beam polarised in the $+z$ direction with an equal one polarised in the $-z$ direction leads to an unpolarised beam.
Now consider the case that the beams have not been mixed {\it physically}, but only {\it mathematically},
and that experiments are performed randomly on either beam.
An observer aware of the non-mixing would still describe them as two separate beam, with measurements giving $s_z=+1$ and  $-1$ in the way he expects;
an observer unaware of the non-mixing  would describe the situation by an unpolarised beam.
Both observers would give a consistent and testable description of the physical situation.} 
So no physical interpretation of the unpolarised beam can be given. This of course changes when a measurement is performed.

 For the description of measurements, however, one obviously wishes to choose the measurement basis as decomposition basis, that is,
 one wishes to interpret in the final density matrix ${\cal D}$ of the combined system at the final time of the measurement, $t_{\rm f}$, 
 in a straightforward manner. Let us consider the example of the Curie-Weiss model.
In the theoretical description of an ideal measurement of a spin $\half$, the outcome for the spin can be $|\hspace{-1mm}\up\rangle$, 
indicated by the apparatus being in a thermodynamic state  with positive magnetisation, described by the mixed \footnote{It  is unrealistic 
that assume that a macroscopic apparatus can be in a pure state, because then a macroscopic number of data should be known; 
still, this is often assumed in measurement theory. } 
density matrix ${\cal R}_\Up$, or the spin can be or $|\hspace{-1mm}\down\rangle$, with the magnetisation in the state ${\cal R}_\Down$, 
with the Born probabilities $p_\up$ or $p_\down$, respectively, 

 \BEQ\label{Dtf=}
 {\cal D}(\tf)=p_\up|\hspace{-1mm}\up\rangle\langle\up\hspace{-1mm}|{\cal R}_\Up+
 p_\down|\hspace{-1mm}\down\rangle\langle\down\hspace{-1mm}|{\cal R}_\Down,
 \EEQ 
This is indeed the result in the Curie-Weiss model after the truncation and registration phases \cite{ABNCW2003,ABNOpus}.
(The off-diagonal terms $\hspace{-1mm}\up\rangle\langle\down\hspace{-1mm}|$ and $\hspace{-1mm}\down\rangle\langle\up\hspace{-1mm}|$ are formally
also present, but their apparatus density operators do not contribute to physical observables, since they are sums over many phase factors.)
It is tempting, common practice and even physically expected, to consider each of the two terms of (1) as describing a subensemble, 
one consisting of events with spin up and pointer variable up, the other one with spin and pointer down. 
However, due to the above quantum oddity, a mixed density matrix such as (\ref{Dtf=}) allows a mathematical decomposition in any basis, typically
having components both on  $|\hspace{-1mm}\up\rangle$ and $|\hspace{-1mm}\down\rangle$. 
May one thus neglect the multitude of other decompositions? Till recently, 
there has been no compelling argument to do this, and hence to explain that (\ref{Dtf=})  has the desired physical interpretation. 
 This conundrum  is called the {\it measurement problem}, as formulated by Lalo\"e \cite{Laloe}~\footnote{Of course, the situation is not so severe that experimentalist 
 are hampered by it. They have long considered the problem as irrelevant in practice.}. 
 
 Considered by some as the deepest problem of quantum mechanics, the impact of the ``unsolvable'' measurement problem has been profound.
It has led to many attempts to find peace with the situation, of which we mention:  
the Copenhagen interpretation, the relative state interpretation  \cite{Everett}, the many worlds interpretation \cite{DeWitt}, the mind-body issue \cite{Wigner},
the modal interpretation \cite{vanFraassen,Dieks}, 
the Bohmian interpretation \cite{Bohm}, the consistent histories interpretation \cite{griffiths_book},
the real ensemble \footnote{In plain terms, one may say that Smolin puts forward that when a certain measurement is done, an ensemble of ``little green men" 
at remote locations in the Universe is doing the same measurement, and employing faster-than-light exchange, they together establish the Born probabilities.} 
interpretation  \cite{Smolin} presented at the 2011 EmQM-1 meeting in Vienna,   
 or the road to extensions of QM, notably spontaneous collapse models \cite{GRW,Pearle}.
Einstein supported the statistical interpretation,  although he never specified its precise meaning. \cite{Fine}

\subsection{Answer to the measurement problem: a physical way out of a mathematical embarrassment}

A resolution based on a new dynamical effect was offered recently within standard quantum statistical mechanics \cite{ABNOpus, ABNOpusculo}:
Near the end of the measurement, realistic, weak dynamical effects inside  the apparatus, that do not change already attained final  the value of the pointer, 
make most decompositions of the density matrix {\it unstable}.
Only the decomposition on the measurement basis is dynamically stable, so this is identified as the physical basis 
towards which the many possible decompositions all relax.
This {\it relaxation} thus indeed allows to interpret  each of the terms of (1) as describing a physical subensemble.
Each of the two thus obtained subensembles being {\it pure} as regarding the tested system, implies that all its members will yield same 
measurement outcome. It is via this pure-state connection that we are allowed to make a transition from an {\it ensemble description}
to {\it individual systems} and {\it individual measurements}.

Arbitrary subensembles can then be decomposed on this basis, having the form (1) with unknown coefficients $q_\up$, $q_\down$ for each subensemble,
constrained only by their sum. \cite{ABNOpusculo}
The obtained structure connects to ordinary probabilities in the frequency interpretation. \cite{vMises}

\subsection{Lessons worth taking}

Looking back we may see what which ingredients were needed to reach the answer.
First, it should not be hard to admit that a quantum measurement involves the interaction between two quantum systems, the tested system and the apparatus. 
Measurements are thus a special type of interacting quantum systems.
\footnote{A broken apparatus may still undergo a similar interaction with the tested system, but it would no longer act as a useful measurement device.}
The apparatus must have some classical features to produce a macroscopic pointer variable, but it is intrinsically quantum as well. 
\footnote{The old idea that the apparatus is classical focusses too much on its pointer variable; apparatuses are made of atoms and hence
intrinsically quantum.}
Realistic modelling of the measurements should acknowledge that;
approaches that try to do away with measurements or apparatuses look devoid of a proper physical setting.
We hold the opinion that the elimination of the apparatus in the theory of measurements
is no less serious than its elimination in the experiment! \cite{ABNOpus}.

Next,  as is fitting with quantum solid state theory,  one adopts the {\it statistical} or {\it ensemble interpretation} of QM \cite{Balian_book}. 
A quantum state is characterised density matrix, which in the limiting case of a pure state is described by a wave function.
Either case, it describes {\it our knowledge} about an ensemble of identically prepared physical systems, \cite{ABNOpus,vanKampen,FuchsPeres}
so pure states have the same meaning as mixed states, they describe also ensembles, be it purified ones.
For the Large Hadron Collider the ensemble could be taken as all collisions within one year of data taking.
A quantum measurement describes an ensemble of physical measurements on the elements of the ensemble of systems.

The statistical interpretation must have been very appealing to the founding fathers, it was mentioned in a book by Kemble in 1937 \cite{Kemble}.
Indeed, it is in spirit close to statistical mechanics, which was well understood when QM was developed. 
Still, subtleties related to the understanding of measurements led the founding fathers to more elaborate constructs, unnecessary as we now have pointed out.
My late teacher Nico van Kampen also strongly supported the statistical interpretation. \cite{vanKampen}\footnote{Van Kampen's Theorem IV reads: 
{\it Whoever endows $\psi$ with more meaning than is needed for computing observable phenomena is responsible for the consequences.} 
}
Arguments against the statistical interpretation based on Bell's theorem \cite{Fine} will be refuted in section 3.2, where we discuss the contextually loophole.

An essential condition  is that the apparatus starts in a {\it metastable state}, 
a realisation of the abstract ``ready state'' often encountered in measurement literature, 
from which  the measurement drives the apparatus into one of the stable states, 
a transition indicated by a change in the pointer. Hence the small quantum signal is amplified by the phase transition, 
as happens visibly, e.g., in  bubble and droplet chambers. After the measurement the apparatus has to be reset in the metastable state.
Mixed initial states avoid the {\it unitarity paradox}. \footnote{The unitarity paradox shows up when the tested system is in a pure state
and one (unphysically) assumes the apparatus to be in a pure state as well. Then the total state is pure, which under the unitary quantum dynamics is 
incompatible with the final state being mixed.}$^,$\footnote{The unitary paradox for a macroscopic black hole is a related misconception,
avoidable by realising that we can never have enough information to describe the black hole by a pure state.}

\section{Elements of a subquantum theory}

We have thus reached the conclusion that the minimal interpretation of QM is the statistical or ensemble interpretation.
The question that immediately comes to the mind is: the ensemble of what? The natural answer is of course: of physical systems.
Hence atoms exist, protons and electrons exist, matter exists, and QM makes statistical statements about measurements on ensembles thereof.

Next question is then: what does an electron look like? One line of research connects it to Kerr-Newman black holes in general relativity.
An intriguing number of phenomena occurring in this classical theory corresponds to notions known from Dirac theory. \cite{Burinskiy}

But first comes the question: what causes quantum behaviour of these quasi-classical objects? 

 \subsection{On the quantum vacuum}
\hfill{ \it You were within }

\hfill{ \it and I sought you outside }

\hfill{St. Augustine}

\vspace{3mm}

Is is known from experiments that the Casimir effect is a real physical effect. There exists a simple analog in ordinary life:
Boats in harbours Òattract each otherÓ, they end up lying in pairs, triples and so on. 
This happens because fewer water surface waves fit between them than on the outside, causing an effective attraction.
Likewise, if we suppose that the quantum vacuum is a real physical vacuum, zero point fluctuations are due to real fields, 
which induce forces on particles, and may, in principle, explain quantum behaviour from classical mechanics.
As a task, this is pretty risky: once the quantum vacuum has been defined, there is a theory with no adjustable parameters, which 
has to explain the very many known quantum effects established  already. But if Nature has this structure, things will just work out, one by one.

While the quantum vacuum would not exist inside its excitations, ``point'' particles, the solitonic nature we presume particles have, 
would imply that they have a finite size, inside which the vacuum fields would likely also pervade.  

Interestingly, quantum-like behaviour is found in the motion of oil drops, which, when tapped, collide with surface waves 
created earlier.  \cite{Couder,BradyAnderson2014}
This {\it proves the possibility} that true quantum behaviour originates from classical stochastic forces.
On another track, De Raedt and Michielsen have designed learning algorithms for event-by-event simulation, 
which reproduce the quantum predictions for many photon and neutron experiments without any quantum input. \cite{deRaedtMichielsen}
Both these subjects motivate to continue the search for a subquantum mechanics.


Concerning the vacuum, a further question is: Up to which energy is it filled? 
A physical picture would be that the vacuum fields gets created soon {\it after}, but not at, the beginning of the Universe.
The maximal filling energy (cutoff energy) would typically be below Planck energy, hence there would be quantum behavior up to that energy scale,
but not at the Planck scale.
Vacuum energy (and pressure) are not expected to come ``out of nothing"', but be borrowed from (compensated by) gravitation.
The cosmological constant could then be protected by energy conservation; still fine tuning seems needed get it to today's small value. 
\cite{NBH11,NEmQM12}

If the quantum vacuum is not filled up to the Planck energy, quantum gravity seems useless, and string theory can only be an effective 
theory. \footnote{String Theory is a framework that may apply to a diversity of problems. Having assumed a large set of (super)symmetries, it possesses 
properties that bewilder its practitioners. However, it does not reproduce the Standard Model of particle physics. Even if it achieves to do so, there will 
remain the question: how to explain this set of fine tuned parameters? Multiverse arguments are equivalent to evading an answer.
Hence string theory cannot be a fundamental theory of Nature.}

\subsection{The need for a subquantum theory}

\hfill{\it A friend in need is a friend indeed}

\hfill{Proverb}
\vspace{3mm}

When in a laboratory a quantum measurement  is performed, we can trust, after nearly a century of tests, 
that the statistics of the outcomes is given by Born's rule. But at the end of one specific measurement we can ask ourselves another question: 
what went on in the apparatus to produce this outcome? And a theorist may ask: with which theory should it be described?
The answer is embarrassing: {\it we have no theory to describe individual events}. \footnote{The approach of \cite{ABNOpus, ABNOpusculo}
answers the measurement problem and thus allows to make the statement that an individual measurement yields one of the quantum outcomes.
One may then hold the opinion that Nature is fundamentally random, and that no more than QM can be known about it.
We do not support this position, but expect that the quantum randomness can be traced back to a stochastic vacuum field.} 
But Nature employs it in every meassurement, so such a theory must exist.
It should be less statistical than QM, but reproduce QM at its statistical level.
 Very  likely it is a hidden-variables theory.

 Contrary to the current general opinion, hidden variables theories are not ruled by Bell's theorem, since it suffers from the {\it contextually loophole},
which can not be closed \cite{TheoBellFoP}.
Indeed, detectors also have hidden variables, so the results from different measurement setups can not be combined in a 
hidden-variables-theory-for all-detector-positions,  so that one can {\it not derive} the Bell inequality. Hence its violations have no physical implications:
absence-of-local-realism arguments based on Bell's theorem should better retire.
The same holds for Bell-type arguments against the statistical interpretation.

 The sought subquantum theory should in any case work for measurement apparatuses, that is, in 
 the GeV scale for the nuclei, the MeV scale for the electrons, the eV energy scale for the atoms and in the meV scale for their functioning. 
 For the GeV and MeV scales fluctuating strong and weak vacuum fields are expected too.
 The Planck scale needs not be involved and space-time can just
 be the good old Minkowski space-time: no extra dimensions, no curvature would be involved at its first level of description.
 On cosmological scales curvature would enter, though, by the fact that matter attracts other matter and light.

\subsection{The classical electron's near field} 

\newcommand{\bm}{{\bf m}}
\newcommand{\bn}{{\bf n}}
\renewcommand{\bn}{\hat{\bf r}}
\newcommand{\ba}{\dot{\bf v}}
\newcommand{\bv}{{\bf v}}

It has been proposed that particles like an electron or a positron, are solitons (solitonic field configurations), 
affected by vacuum fluctuations that act on them as stochastic forces. \cite{NRelH,NBtQ,NEmQM12}
Hence a {\it Stochastic Soliton Mechanics} might  underly quantum mechanics \cite{NRelH}, being a less statistical theory, 
in the way that a given realisation of  the Langevin equation is less statistical than the Fokker-Planck equation.

The electromagnetic field of a classical point particle (modelling the exterior of a proton, neutron, electron, positron or an ion)
with charge $q=Ze$, $Z=0,\pm1,\pm2,\cdots$,  and  magnetic dipole moment $\bm$, moving with non-relativistic speed $\bv$, reads

\BEQ
\vE=\frac{q}{4\pi\epsilon_0}\frac{\bn}{r^2}
+\frac{q}{4\pi \epsilon_0c^2}\frac{\bn\times(\bn\times\dot\bv)}{r},\quad
\vB=\frac{\mu_0}{4\pi} \frac{3\bn(\bn\cdot\bm)-\bm}{r^3}
+\frac{\mu_0q}{4\pi} \frac{\bv\times\bn}{r^2}-\frac{\mu_0q}{4\pi c} \frac{\bn\times\dot\bv}{r},
\EEQ
where the first terms are the charge, magnetic dipole and Lorentz  terms, and the last terms the radiation fields.
Clearly, the charge and magnetic dipole terms act as a polarisation of the vacuum, which, as for polarons, travels with the electron. \footnote{Since 
the classical electron radius $\sim1$ fm  is very small, it likely plays no role  in atoms and molecules.}
Since $\bm=g(q/2m) {\bf S}$,  the magnetic dipole field carries, in cases with $|g|={\cal O}(1)$, information about the particle  spin ${\bf S}$.
From their ubiquitous role in quantum theory, one expects a major role for this magnetic dipole term in the cause of spin-spin interactions.
It is thus conceivable that the magnetic dipole moment plays a role in the EPR paradox, when a pair of electrons or an electron-positron pair
separates, with each partner carrying the information about its spin with it in its surrounding $\vB$ field-cloud.
Hence no ``action-at-a-distance'' is needed to explain a correlation between spins of the partners, it will be created at the source
and just be conserved in the motion of the members of the pair.

The question why neutrons do interact with normal matter, while neutrinos hardly do, gets a simple answer as well.
First, from the Compton radius $\hbar/mc$ = 0.21 fm for neutrons while equal to 0.13 $\mu$ for the case that the neutrino has
a mass of 1.5 eV \cite{NhNeutrinoEPL}, one might be tempted to expect that the neutrinos interact much stronger than neutrons. 
This is obviously not the case, so the Compton size is not the important aspect. In contrast, even though both have spin $S=\half$, 
the  absence of a neutrino magnetic moment  (it is often estimated to be $10^{-19}\mu_B$), 
directly explains their inertness in matter compered to neutrons with their $g_\mathrm{n}= -3.8$.
\footnote{Neutrons have $g_\mathrm{n}= 	-3.82608545\pm 0.00000090$, while for protons $g_\mathrm{p} = 5.585694713 \pm	0.000000046$.}

 \section{Stochastic electrodynamics and the hydrogen problem}

Stochastic Electrodynamics (SED) is a theory that views the world as classical, with a real vacuum field composed of fluctuating electrodynamics fields. 
Each mode of the vacuum with frequency  $\om=ck$ is supposed
to have the zero-point Planck energy $\half\hbar\om$, implying a spectrum

\BEQ
\rho_{ZP}(\om)\d\om=2\frac{4\pi k^2\d k}{(2\pi)^3}\half\hbar\om=\frac{\hbar\om^3}{2\pi^2c^3}\d\om,
\EEQ
where the factor 2 arises from the polarisation modes.
This $\om^3$ spectrum is expected to hold up to a very large cutoff $\om_c=E_c/\hbar$.

The theory is known to work well for harmonic oscillator problems, while generally suspected to fail beyond.
However, a general structure for `nonlinear' problems has been put forward in the books by 
de la Pe\~na, Cetto and Vald\'es.  \cite{delaPenaCettobook,delaPenaCettoValdesbook}

Let us thus consider the most known, nonlinear problem: the hydrogen atom in its ground state. 
Within SED this is the motion of a classical point charge $-e$ in the field of a central charge $e$.
The Newton equation with damping and noise, also called Abraham-Lorentz equation or Brafford-Marshall equation,
 reads

\BEQ \label{eeqnmot}
m\ddot\vr=-\frac{e^2\vr}{4\pi\epsilon_0r^3}
+\frac{e^2}{6\pi\epsilon_0c^3}\,\dot{\ddot {\vr}}-e\vE(\vr,t) 
\qquad \textrm{\bf How to implement the latex ``dddot'' for } \d^3\vr/\d t^3 ??
\EEQ
The stochastic fields are described by a vector potential
\BEQ
\vA&=&
\sum_{n_1n_2n_3\lambda}
\sqrt{\frac{{\mu_0{\cal E}_\vn}}{V}}
\hat\eps_{\vn\lambda}[A_{\vn\lambda}\sin(\vk_\vn\cdot\vr-\om_\vn t)
+B_{\vn\lambda}\cos(\vk_\vn\cdot\vr-\om_\vn t)] ,
\EEQ
where the $\hat\eps_{\vn\lambda}$ with $\lambda=1,2$ are polarisation vectors, 
the $A_{\vn\lambda}$ and $B_{\vn\lambda}$ are independent Gaussian random amplitudes with average 0 and variance 1
of the plane waves labeled by $\vk_\vn=2\pi \vn/V^{1/3}$, with $\vn=(n_1,n_2,n_3)$ having integer components.
This determines the electric field $\vE=-\dot\vA$ and magnetic field $\vB=\nabla\times\vA$, which together lead to an energy $\half\hbar\omega$ per mode
in the volume $V$.
With Bohr frequency $\om_0$ defined by twice the Rydberg energy,  $\hbar\om_0\equiv 2R_y=\alpha^2mc^2$, 
where $\alpha=e^2/4\pi\epsilon_0\hbar c\approx 1/137$ is the fine structure constant, 
and the Bohr radius $a_0=\hbar/\alpha mc$, it is seen that
 for frequencies $\om\sim \om_0$, the wavelength is of order $a_0/\alpha$, implying that the field is uniform over the extension of the atom.
 Hence one can adopt the dipole approximation $\vk\cdot\vr\to0$ in the $\vE$ and $\vB$ fields, producing

\BEQ
\vE&=&
\sum_{n_1n_2n_3\lambda} \sqrt{\frac{{{\cal E}_\vn}}{\epsilon_0V}}
\hat\eps_{\vn\lambda}[A_{\vn\lambda}\cos\om_\vn t+
B_{\vn\lambda}\sin\om_\vn t], \EEQ
and a similar expression for $\vB$. To leading order, the latter is not needed, since in (\ref{eeqnmot}) we could omit the Lorentz force
$-e\,\vv\times\vB$, since it is smaller than the electric force $-e\vE$ by a factor
$v/c\sim \om_0a_0/c=\alpha$, and we shall only be interested in the leading order in $\alpha$.

\myskip{ 
 \vB&=&
\sum_{n_1n_2n_3\lambda}
\sqrt{\frac{{\mu_0{\cal E}_\vn}}{V}}\hat k_\vn\times
\hat\eps_{\vn\lambda}[A_{\vn\lambda}\cos \om_\vn t
+B_{\vn\lambda}\sin\om_\vn t]\EEQ 
}
\myskip{\BEQ
 \vE&=&
\sum_{n_1n_2n_3\lambda} \sqrt{\frac{{{\cal E}_\vn}}{\epsilon_0V}}
\hat\eps_{\vn\lambda}[A_{\vn\lambda}\cos(\vk_\vn\cdot\vr-\om_\vn t)
-B_{\vn\lambda}\sin(\vk_\vn\cdot\vr-\om_\vn t)],
\nn\\
 \vB&=&
\sum_{n_1n_2n_3\lambda}
\sqrt{\frac{{\mu_0{\cal E}_\vn}}{V}}\hat k_\vn\times
\hat\eps_{\vn\lambda}[A_{\vn\lambda}\cos(\vk_\vn\cdot\vr-\om_\vn t)
-B_{\vn\lambda}\sin(\vk_\vn\cdot\vr-\om_\vn t)]\EEQ 

The energy of these modes in the volume $V$ is indeed

\BEQ \label{zpen}\frac{1}{2}\int _V\d^3r(\epsilon_0\vE^2+\frac{\vB^2}{\mu_0})=\sum_{\vn\lambda}{\cal E}_{\vn}
\frac{A_{\vn\lambda}^2+B_{\vn\lambda}^2}{2}=\sum_{\vn\lambda}{\cal E}_\vn,\qquad \frac{1}{\mu_0}=c^2\epsilon_0,\qquad \mu_0\to4\pi c^2
\EEQ
}

With the equation of motion (\ref{eeqnmot}) for the electron and its stochastic field fixed statistcially, the first question is whether 
it will produce a stable ground state with the statistical properties known from quantum mechanics. 
In absence of direct analytical methods, this is a problem for simulations.
An encouraging answer was given in 2003 by Cole and Zou \cite{ColeZou2003}. 
Further investigations are being carried out at the University of Amsterdam \cite{NLvH2014}.

\section{A subquantum arrow of time}



\hfill{\it Time will consume all things including itself}

\hfill{ Indian proverb}

\vspace{3mm}

We now have finished the preparations to arrive at the main theme of this work.
Though motivated by SED we now focus on a broader class of similar subquantum theories, like \cite{Khrennikov,Groessing}.

If the equation of motion  (\ref{eeqnmot}) has a stationary distribution, it will be a state with energy output by the radiation term
and energy input from the average behaviour of the stochastic electric field. Hence there is a current of energy involved, an 
{\it energy throughput}, which means that there is an {arrow of time}. We call this the {\it subquantum arrow of time}.
It goes hand-in-hand with the stability of atoms and matter.

This subquantum arrow of time is more fundamental than the entropic arrow of time, since it holds already for each of the involved atoms.
It is likely more fundamental than the cosmological arrow of time, since one expects that the expansion of the Universe
has only minuscule impact on laboratory physics. The quantum arrow of time related to wave function collapse
in a measurement is related to irreversibility in the measurement apparatus \cite{ABNOpus}, so it is akin to the thermodynamic arrow of time
and also less fundamental than the subquantum arrow of time.

\section{Summary}

{} From the analysis of ideal measurements, we conclude that QM itself describes the statistics of measurement outcomes.
The minimal interpretation is the statistical one; other interpretations involve additional assumptions of one type or another, which can be
omitted by Occam's razor.
To understand ideal measurements
{\it no measurement postulates are needed} and {\it no extension of QM is needed}, which is a very satisfactory solution to this century-long problem. 
There is {\it no role for the observer} (which can be an automated routine such as used at the LHC to pre-select desired events), 
except that observation allows to update the knowledge about the ensemble  (``wave function collapse'').
The occurrence of {\it individual events} and their classical probability structure is  {\it described by quantum theory} as well, due to a 
new relaxation mechanism which takes place inside the apparatus near the end of the measurement.

We advocate the statistical formulation for the teaching of quantum theory, since it fully explains ideal measurements. 
The concept of a quantum state is simple to grasp by being in spirit close to states in classical statistical physics.
States described by wave functions should be regarded only as special cases, since pure states as well as mixed states describe (sub)ensembles.

Non intuitive features of quantum mechanics remain concentrated in the non commutation of the observables representing the physical quantities. 
The situation may be further specified in a subquantum mechanics, and in particular in Stochastic Electrodynamics.
In such theories the atomic stability will be induced by a classical vacuum, which pervades all space, inside and outside atoms. 
It has probably been created very early in the Universe, but possibly well after the Planck time.
This vacuum enables the stability of matter by a throughput of energy, hence it carries a ``subquantum'' arrow of time, combining the stability of matter with the 
forward direction of time. The questions ``why does matter exist?'' and ``why is time increasing?''  may thus appear to be intimately  connected.

\end{document}